\documentclass{article}
\usepackage{emulateapj}
\usepackage{apjfonts}

\begin{document}

\submitted{To appear in The Astrophysical Journal Letters}

\title{Isolating Clusters with Wolf-Rayet Stars in I Zw 18$^1$}


\author{Thomas M. Brown}

\affil{Space Telescope Science Institute, 3700 San Martin Drive,
Baltimore, MD 21218.  tbrown@stsci.edu}

\author{Sara R. Heap, Ivan Hubeny, Thierry Lanz, Don Lindler}

\affil{Code 681, NASA Goddard Space Flight Center, Greenbelt, MD 20771. 
heap@srh.gsfc.nasa.gov, hubeny@tlusty.gsfc.nasa.gov, lanz@nova.gsfc.nasa.gov,
lindler@rockit.gsfc.nasa.gov}

\begin{abstract}

We present UV images and spectra of the starburst galaxy I Zw
18, taken with the Space Telescope Imaging Spectrograph.  The high
spatial resolution of these data allows us to isolate clusters
containing Wolf-Rayet stars of the subtype WC.  Our far-UV spectra
clearly show \ion{C}{4} $\lambda\lambda$1548,1551 and \ion{He}{2}
$\lambda$1640 emission of WC stars in two clusters: one within the
bright (NW) half of I Zw 18, and one on the outskirts of this region.
The latter spectrum is unusual, because the \ion{C}{4} is seen only in
emission, indicating a spectrum dominated by WC stars.  These data
also demonstrate that the \ion{H}{1} column in I Zw 18 is strongly
peaked in the fainter (SE) half of I Zw 18, with a column depth far
larger than that reported in previous analyses.

\end{abstract}


\section{INTRODUCTION}

I Zw 18 (Mrk 116) represents an important local fiducial in our
understanding of galaxy formation and evolution.  It is a blue compact
galaxy with the smallest known abundance of heavy elements in its
ionized gas ($\sim$2\% of solar; V$\acute{\rm i}$lchez \&
Isglesias-P$\acute{\rm a}$ramo 1998 and references therein).  For the
past three decades, the debate has focused on the age of its stellar
populations.  Early analyses of Wide Field Planetary Camera 2 (WFPC2)
images suggested that star formation began several tens of Myr ago
(Dufour et al. 1996; Hunter \& Thronson 1995), while later analyses of
the same WFPC2 images (Aloisi, Tosi, \& Greggio et al.\ 1999) and new
images from the Near Infrared Camera and Multiobject Spectrometer
($\ddot{\rm O}$stlin 2000) implied the presence of an older population
(0.5--1 Gyr) from a small number of faint red stars.

Two independent sets of ground-based observations have recently
discovered Wolf-Rayet (WR) emission within the brighter (NW) component of I
Zw 18: Izotov et al.\ (1997) found evidence for 17 WN and 5 WC stars,
while Legrand et al.\ (1997) found evidence for 1--2 WC stars.  While
neither group had the spatial information to tie this emission to
stellar clusters, the existence of WR stars in I Zw 18, especially of
subtype WC, is surprising, because canonical (non-rotating and
non-binary) evolutionary theory predicts few WN stars and no WC stars
to form at the low metallicity of I Zw 18 (Cervi$\tilde{\rm n}$o \&
Mas-Hesse 1994; Meynet 1995).  de Mello et al.\ (1998) subsequently
searched for these WR stars in narrow-band WFPC2 images, and found
faint peaks in the \ion{He}{2} $\lambda$4686 emission they attributed
to either nebular emission or 5--9 WN stars.

Using the Space Telescope Imaging Spectrograph (STIS) on board the
Hubble Space Telescope (HST), we recently obtained a two-dimensional
far-UV spectroscopic map of I Zw 18, along with far-UV and near-UV
images, in \linebreak

{\noindent $^1$Based on observations made with the NASA/ESA Hubble Space
Telescope, obtained at the Space Telescope Science Institute, which is
operated by the Association of Universities for Research in Astronomy,
Inc., under NASA contract NAS 5-26555. These observations are
associated with proposal 9054.}

\noindent 
order to further understand its star-formation history.  These data
reveal two clusters containing WC stars.  While neither of these
clusters is associated with the WR detections of Legrand et al.\
(1997) and de Mello et al.\ (1998), one of the clusters might be
responsible for the WC emission seen by Izotov et al.\ (1997).  Our
STIS data also provide a high-resolution \ion{H}{1} map, based on the
Ly-$\alpha$ absorption; we show that the \ion{H}{1} column depth
($N_{HI}$) is sharply peaked within the SE component.

\section{OBSERVATIONS}

We obtained far-UV spectra of I Zw 18 from 2002 Feb 9 to 2002 Feb 12,
using the G140L grating and the $52\arcsec \times 0.5\arcsec$ slit on
STIS (for a full description of the instrument, see Woodgate et al.\
1998 and Kimble et al.\ 1998).  The field-of-view of the STIS UV
detectors is $25\arcsec \times 25\arcsec$.  We placed the slit at
seven adjacent positions in I Zw 18, parallel to the axis of the main
body (position angle = 145$^{\rm o}$), resulting in spatially resolved
spectra of its bright star-forming regions (Figure 1).  The exposure
time at four positions was 6058 s (the central 3 positions and the SW
extreme), while at the remaining three positions it was 5376 s.  At
each slit position, half the exposures were offset from the others by
$0.5\arcsec$ (20 pixels) along the slit, in order to allow masking of
bad, hot, and shadowed pixels in the coaddition, and to smooth over
variations in the detector response.  The spectra were processed
through the standard calibration pipeline, except that we also
subtracted an appropriately scaled profile of the dark rate, from a
sum of 500 ks of dark exposures.  The subtraction of this dark
``glow'' (which peaks in the upper left-hand quadrant and can be 20
times higher than the nominal $7\times 10^{-6}$ cts s$^{-1}$
pix$^{-1}$ dark rate) is important when studying faint extended
sources.  Finally, we subtracted the airglow spectrum using a region
in each spectral image free from source flux.

On 2002 Feb 25, we also obtained far-UV and near-UV images of I Zw 18
using the F25SRF2 and F25QTZ filters, respectively, with bandpasses
centered at 1457~\AA\ and 2365~\AA, and exposure times of 5331 s and
5786 s.  We again used $0.5\arcsec$ offsets to allow masking of pixels
and smoothing of detector response.  All of the frames for each camera
were coadded, rescaled, and shifted with the DRIZZLE package in IRAF,
in order to match the G140L spectral images described above.

In addition to our STIS data, we also obtained archival WFPC2 images
at longer wavelengths, in the F555W ($V$), F439W ($B$), F336W, and
F469N (\ion{He}{2} $\lambda 4686$) filters, for comparison with
earlier work.  The WFPC2 data were cleaned of cosmic rays, coadded,
and drizzled to register with the STIS data.  Figure 1 shows a false
color image produced from our STIS data in the blue and green
channels, and the WFPC2/F555W data in the red channel.

\vskip 0.1in
\parbox{7.0in}{\epsfxsize=7.0in \epsfbox{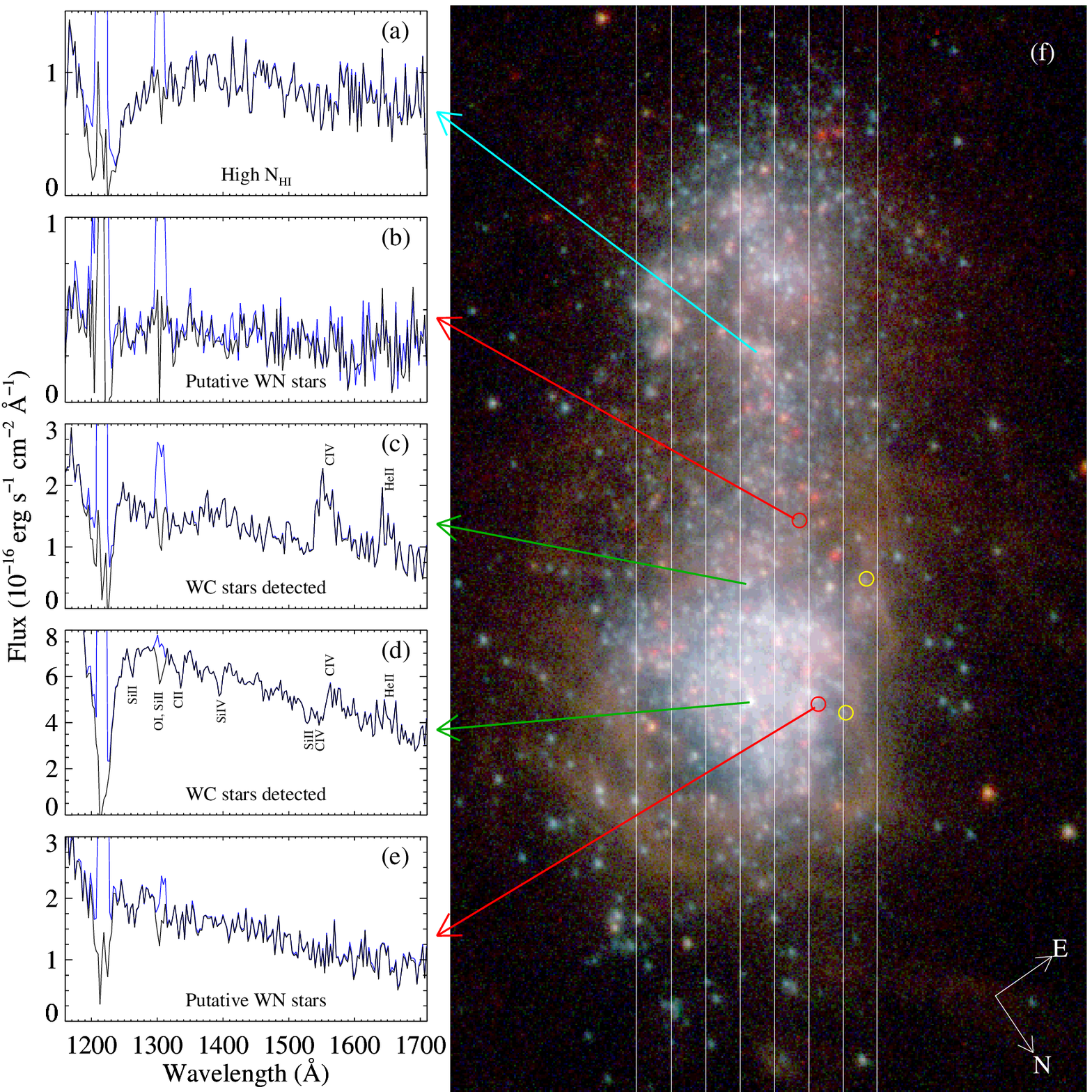}}

\vskip 0.1in
\parbox{7.0in}{\small {\sc Fig.~1--} Far-UV spectra and broad-band
images of I Zw 18.  Panel a: The spectrum with the broadest
Ly-$\alpha$ absorption and thus the highest $N_{HI}$, with
(black) and without (blue) airglow subtraction.  Panels b and e:
Spectra of regions with previously reported WN detections (de Mello et
al.\ 1998); we find no evidence for WN stars.  Panels c and d: Spectra
of clusters with WC stars; note the \ion{C}{4} and \ion{He}{2} emission
lines.  Panel f: Composite HST image of I Zw 18 in 3 bandpasses
(blue = STIS/FUV/F25SRF2, green = STIS/NUV/F25QTZ, red = WFPC2/F555W).
The seven adjacent positions of the STIS slit are marked (white
lines).  Many of the old stars reported by Aloisi et al.\ (1999)
appear as faint red stars here, but some are UV-bright, indicating a
hotter temperature or a hot companion.  Yellow circles mark two regions
where we find possible narrow \ion{He}{2} emission but no associated
\ion{C}{4} emission. The de Mello et al.\ (1998) WN detections are
encircled in red.}

\noindent

\section{ANALYSIS}

\subsection{WC Stars}

The CALSTIS pipeline software produces two-dimensional rectified
spectral images that are useful for examining the spectral energy
distributions (SEDs) of extended objects.  These images are linear in
wavelength along the x-axis, and linear in spatial extent along the
y-axis.  Upon examination of these images, we saw a compact source
with obvious WR/WC emission \linebreak

\newpage

\noindent
features at \ion{C}{4} $\lambda\lambda
1548,1551$ and \ion{He}{2} $\lambda 1640$ (shown in Figure 1$c$).  We
then performed a systematic search for other such objects in the data.

To find the locations of individual spectra in these images, we
collapsed the spectra along the dispersion axis and convolved with a
Gaussian normalized to unit area and zero sum (similar to the finding
algorithm of the DAOPHOT package; Stetson 1987).  The local maxima in
these convolutions gave the y-positions of compact sources along the
slit.  Returning to the original spectral images, we extracted spectra
at each of these positions, using row sums 5 pixels high.  We also
extracted spectra from the two regions with significant WR/WN
detections in previous analyses of WFPC2 data (de Mello et al.\ 1998; those not
marked with ``?'' in their Table 3), although these regions are not
associated with compact sources in our spectral images, or with
sources in the UV and optical HST images (compare their Figures 2 and
3$a$).  Each spectrum was then shifted to account for the position of
the object within the slit, using a centroid on the corresponding
region in the far-UV image.

After extracting 205 spectra at the local maxima in our spectral
images, we performed a systematic search for broad \ion{He}{2}
$\lambda 1640$ emission, using as a guide the spectrum of the source
with obvious WR/WC features discussed above.  We fit a linear
continuum to each spectrum, using spectral regions free of strong
emission and absorption features (1400--1500~\AA, 1590--1620~\AA, and
1670--1710~\AA), summed the net flux exceeding this continuum in the
region 1635--1665~\AA\ (the heliocentric velocity of I Zw 18 is 751 km
s$^{-1}$; Thuan et al.\ 1999), and returned a list of $> 3 \sigma$
detections of excess flux near 1640~\AA.  The automated search
detected the obvious source we had already found by eye, plus one
additional source of WR emission; both spectra also show broad
\ion{C}{4} emission, indicating the presence of WC stars (Figure 1,
panels $c$ and $d$).  Note that the emission lines are much broader
than a resolution element in our spectra ($\sim 1$~\AA\ for point
sources and $\sim 12$~\AA\ for sources filling the slit).  We also
found two spectra with narrow emission features near 1640~\AA\ that
could be noise or nebular emission; the spectra do not have \ion{C}{4}
emission, and their associated clusters are marked in Figure 1$f$ (yellow
circles).

Our WC detections are located $\sim1.3\arcsec$ SE and $\sim
0.5\arcsec$ W of the center of the NW starforming region.  Legrand et
al.\ (1997) report WC emission $\sim 1.5\arcsec$ SW of the central
cluster, so our detections are distinct.  Izotov et al.\ (1997) only
restrict their detections of WR stars to the NW starforming region.
None of the detections of WN stars by de Mello et al.\ (1998)
correspond to the positions of our WC stars; in Figure 1$b$ and 1$e$,
we show spectra from the regions of their putative detections.

To estimate the number of WR/WC stars needed to produce the observed
\ion{He}{2} emission, we used the tabulations of Schaerer \& Vacca
(1998).  In the middle of the range of WC stars, WC5 stars produce an
average of \ion{He}{2} luminosity of $5\times 10^{36}$ erg s$^{-1}$.
Assuming a distance of 12.6 Mpc ($\ddot{\rm O}$stlin 2000) and a
foreground extinction of $E(B-V) = 0.03$~mag (Schlegel, Finkbeiner, \&
Davis 1998) but no reddening from the patchy internal dust (Cannon et
al.\ 2002), we can convert our \ion{He}{2} fluxes to luminosity.  For
the spectra shown in Figure 1$c$ and 1$d$, we measure $L_{1640} =
(2.4\pm 0.5) \times 10^{37}$ erg s$^{-1}$ and $L_{1640} = (3.2\pm 1.0)
\times 10^{37}$ erg s$^{-1}$, respectively, corresponding to 5$\pm 1$
and 6$\pm 2$ WR/WC stars.  Using the 1500~\AA\ continuum flux and a
5~Myr isochrone (Bertelli et al.\ 1994), we estimate that these
clusters contain 26 and 119 stars, respectively, with with $M >
$8~$M_\odot$, suggesting that the ratio WR/(WR+O) in these clusters is
respectively 0.2 and 0.05.  The stronger WR/WC detection in Figure
1$c$ is due to the higher fraction of WR/WC stars in that cluster;
note the prominent \ion{C}{4} and \ion{Si}{4} absorption in Figure
1$d$, where the hot underlying population of O stars is more dominant.
Our measured WR/(WR+O) ratios fall within the range observed in other
galaxies, but are unusually high given the metallicity of I Zw 18 (see
Maeder \& Meynet 1994).  Of course, the ratio is far lower for the
galaxy as a whole, given the many UV-bright clusters showing no WR
emission.

The ratio of \ion{C}{4} to \ion{He}{2} emission in Figure 1$c$ is
$\sim$2.3, which is typical for the WC subtype (Niedzielski \&
Rochowicz 1994; Gr$\ddot{\rm a}$fener, Koesterke, \& Hamann 2002).
However, this ratio is only $\sim 0.7$ in Figure 1$d$; the smaller
ratio is partly due to the \ion{C}{4} absorption from the
hot stellar population and interstellar medium, but it could also
indicate that one or two of the WR stars in this cluster are of type WN
instead of WC.

de Mello et al.\ (1998) claimed detections of 1.5 and 3 WNL stars,
respectively, where we are showing the spectra in Figure 1$b$ and
1$e$.  Assuming $L_{1640} = 1.6 \times 10^{36}$ erg s$^{-1}$ (Schaerer
\& Vacca 1996), individual WNL stars would be difficult to detect in
our data, but we do not see any signs of these stars.  Where de Mello
et al.\ (1998) found 1.5 WNL and 3 WNL stars, we formally find $-0.5
\pm 3$ WNL stars and $+0.5 \pm 1.6$ WNL stars, respectively.  Note
that there are no point sources in our UV images (or in the extant
WFPC2 images) at the locations of the de Mello et al.\ 
detections, and de Mello et al.\ note that their detections are
consistent with both nebular or WR emission.  Unfortunately, their
F469N image has a hot pixel exactly where we have our strongest
detection of WR/WC emission, which probably precluded their detection
of these sources.

\subsection{Old Clusters or Large Interstellar Hydrogen Column?}

Compared to the spectra in the NW half of I Zw 18, our spectra in the
SE half generally show broader Ly-$\alpha$ $\lambda$1216 absorption,
but it is especially broad in one particular spectrum (see Figure
1$a$).  Without other data, the absorption could arise from two
situations: either a high column density of interstellar \ion{H}{1},
or a predominately older ($ \gtrsim 500$ Myr) and cooler population of
stars.  We investigated this latter possibility by examining the SED
with a longer wavelength baseline than that available in our far-UV
spectra.  Note that if the spectrum in Figure 1$a$ came from a 500 Myr
population, it would be have to be $\sim$60 times more massive than
the cluster shown Figure 1$d$.

Using the HST images in the far-UV, near-UV, $B$, and $V$ bands, we
measured the flux in the region of our spectral extraction (an area
$0.5\arcsec \times 0.12\arcsec$).  This flux comes from a string of
compact sources aligned with the long axis of the extraction box, so
we estimated the aperture correction required by using TinyTim models
of the HST PSF in the different bandpasses.  We do not expect perfect
agreement with the G140L spectrum (due to wavelength-dependent
aperture correction uncertainties in the images and spectrum), but
they agree at the $\sim 10$\% level (see Figure 2).  Furthermore, it
is difficult to quantify exactly the dominant effective temperature of
the stellar population, given the reddening uncertainties and the fact
that we are looking in the Rayleigh-Jeans tail.  However, the SED of
this region is clearly dominated by a population of hot stars.  We
show this by integrating the synthetic spectra of Kurucz (1993) over
the isochrones of Bertelli et al.\ (1994), for two different ages
(Figure 2).  A population of about 500 Myr (or a single star of $\rm
T_{eff}\approx 12,000$~K) is needed to explain the width of
Ly-$\alpha$ absorption without a significant contribution from
interstellar \ion{H}{1}.  However, such a population would produce far
more flux at longer wavelengths than is actually seen.  Instead, a
much younger population (e.g., 16~Myr) reddened by $E(B-V)=0.06$ mag
(SMC extinction law; Witt \& Gordon 2000) can reproduce the spectral
energy distribution from the far-UV to the optical, but this requires
that the Ly-$\alpha$ absorption come from a high \ion{H}{1} column:
$N_{HI} \approx 2 \times 10^{22}$ atoms cm$^{-2}$.  This result is
independent of the dominant $\rm T_{eff}$ in the underlying
population, for $\rm T_{eff} \gtrsim 15,000$~K.  The spectra show much
narrower Ly-$\alpha$ absorption as one moves away from this position;
e.g., looking along a sightline to a bright cluster 1$\arcsec$ to the
SE, the \ion{H}{1} column must be at least ten times smaller (with the
exact amount depending upon the assumed age/temperature of the
underlying population).

van Zee et al.\ (1998) produced high spatial resolution (5$\arcsec$)
maps of the \ion{H}{1} distribution in I Zw 18, using the Very Large
Array (VLA), and found that the column density peaks in the SE half of
the galaxy at $3 \times 10^{21}$ atoms cm$^{-2}$.  Because each of the
two starbursting regions (NW and SE) subtends $\sim 3\arcsec$ on the
sky, a sharp peak in the \ion{H}{1} distribution would be much smaller
than a VLA resolution element, and thus the column density could peak
at a significantly larger value than the peak in the van Zee et al.\
(1998) map.

\subsection{Summary}

We have presented high spatial resolution images and spectra of
the starburst galaxy I Zw 18.  These spectra show strong WR/WC
emission features originating in two compact clusters.  From the
\ion{He}{2} $\lambda$1640 emission, we estimate that 6 WR/WC stars
reside in a very UV-bright cluster situated well within the NW
starbursting region, and that 5 WR/WC stars lie in a somewhat fainter
cluster just outside of this region.  The fact that any WR/WC stars
exist at all in I Zw 18 is surprising, because canonical theory for
the evolution of single, massive, non-rotating stars predicts few WR
stars and no WC stars to form at the low metallicity of I Zw 18
(Meynet 1995; Cervi$\tilde{\rm n}$o \& Mas-Hesse 1994).  

The existence of WR/WC stars at the low metallicity of I~Zw~18
demonstrates the importance of sophisticated models of massive star
evolution.  Maeder \& Meynet (2001) find that the inclusion of
rotation in their models decreases the minimum mass for a single star
to become a WR star, allowing the formation of WR stars at lower
metallicity.  Furthermore, there is both theoretical and observational
evidence for the importance of close binary evolution
in the formation of WR stars,
especially as one moves to lower metallicity (see Bartzakos,
Moffat, \& Niemela 2001); the role of binaries in WR formation has
a long history (see Paczy$\acute{\rm n}$ski 1967), even though much
of the WR literature focuses on single-star evolution.

Our data also show a sharp peak in the \ion{H}{1} column density of I
Zw 18, situated within the SE half of the galaxy.  This peak falls
within the peak of previous high spatial resolution maps of the
\ion{H}{1} taken with the VLA (van Zee et al.\ 1998).  However, our
measurement of 2$\times 10^{22}$ atoms cm$^{-2}$ exceeds the earlier
estimates by an order of magnitude, because of the higher
resolution in our HST data.  

\acknowledgements

Support for proposal 9054 was provided by NASA through a grant from
the Space Telescope Science Institute, which is operated by the
Association of Universities for Research in Astronomy, Inc., under
NASA contract NAS 5-26555.  The authors are grateful to C. Leitherer
for advice and useful discussions.  A. Aloisi kindly provided her 
catalog of WFPC2 photometry.

\noindent
\parbox{3.25in}{\epsfxsize=3.25in \epsfbox{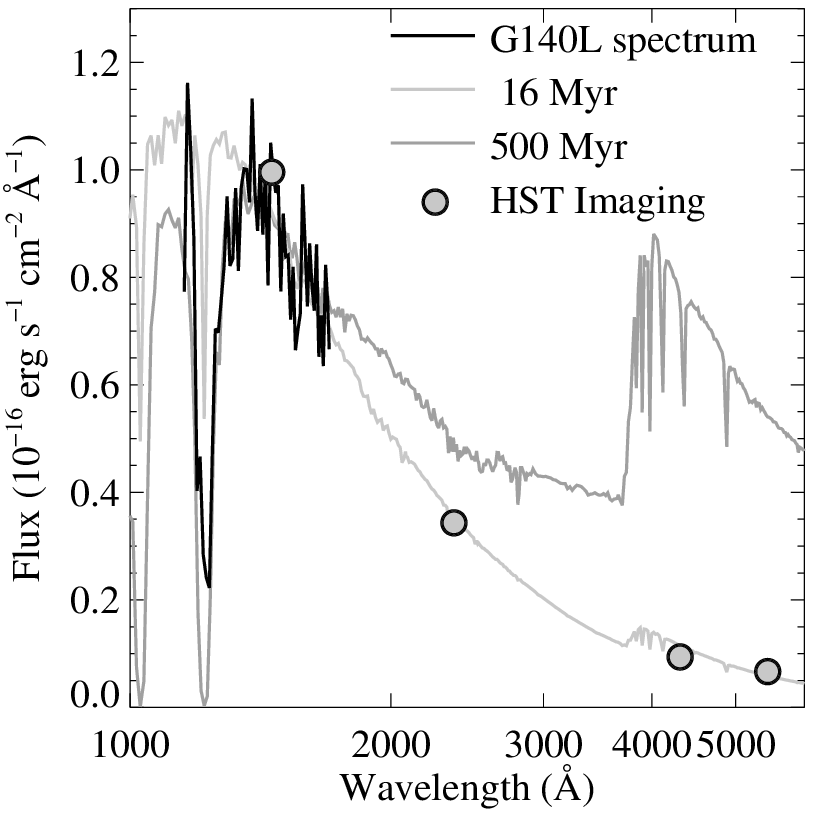}} \\
\noindent
\parbox{3.25in}{ \small {\sc Fig.~2--} Our spectrum showing
the broadest Ly-$\alpha$ absorption (see Figure 1$a$), here over a
longer wavelength baseline, combining the G140L data (black curve) and
HST images (circles).  The flux decline at longer
wavelengths and the small Balmer jump show clearly that this
region is dominated by a population of hot stars.  To
demonstrate, we show the integrated flux from a lightly reddened 16
Myr population (light gray curve) and the flux from an unreddened 500
Myr population (dark gray curve).  An \ion{H}{1} column of $2 \times
10^{22}$ atoms cm$^{-2}$ is required (not shown for clarity) to
produce the width of the Ly-$\alpha$ absorption if a young (hot) population of
stars is assumed.  }

\end{document}